\documentclass[english]{article}
\usepackage[T1]{fontenc}
\usepackage[latin1]{inputenc}
\usepackage{geometry}
\usepackage{color}
\usepackage{amssymb}

\makeatletter

\AtBeginDocument{
  
}

\usepackage{babel}
\makeatother
\begin{document}
\begin{center}\textbf{\large Higher order antibunching is not a rare
phenomenon}{\LARGE }\\
{\LARGE ~}\\
\end{center}{\LARGE \par}

\begin{center}\textcolor{black}{\large Prakash Gupta}%
\footnote{sai\_prakash1313@yahoo.com%
}\textcolor{black}{\large , Pratap Narayan Pandey}%
\footnote{pandey\_jiit@yahoo.co.in%
} \textcolor{black}{\large and Anirban Pathak}\textcolor{black}{}%
\footnote{\textcolor{black}{anirbanpathak@yahoo.co.in}%
}\end{center}

\begin{center}\textcolor{black}{Department of Physics, JIIT, A-10,
Sectror-62, Noida, UP-201307, India.}\\
\end{center}

\begin{abstract}
\textcolor{black}{\normalsize Since the introduction of higher order
nonclassical effects, higher order squeezing has been reported in
a number of different physical systems but higher order antibunching
is predicted only in three particular cases. In the present work,
we have shown that the higher order antibunching is not a rare phenomenon
rather it can be seen in many simple optical processes. To establish
our claim, we have shown it in six wave mixing process, four wave
mixing process and in second harmonic generation process. }{\normalsize \par}
\end{abstract}

\section{\textcolor{black}{\normalsize Introduction}}

\textcolor{black}{Antibunching and squeezing do not have any classical
analogue {[}\ref{elements of quantum optics}-\ref{hbt}{]}. Higher
order extensions of these nonclassical states have been introduced
in recent past {[}\ref{hong1}-\ref{lee2}{]}.  Among these higher
order nonclassical effects, higher order squeezing has already been
studied in detail {[}\ref{hong1}, \ref{hong2}, \ref{hillery}, \ref{giri}{]}
but the higher order antibunching (HOA) is not yet studied rigourously.
The idea of HOA was introduced by Lee in a pioneering paper {[}\ref{lee1}{]}
in 1990, since then it is predicted in two photon coherent state {[}\ref{lee1}{]},
trio coherent state {[}\ref{ba an}{]} and in the interaction of intense
laser beam with an inversion symmetric third order nonlinear medium
{[}\ref{the:martin}{]}. From the fact that in last 15 years HOA is
reported only in three particular cases, HOA appears to be a very
rare phenomenon. The present study aims to establish that this apparent
rarity is not due to any physical reason. To establish that, we have
shown the existence of HOA in six wave mixing process, four wave mixing
process and in second harmonic generation process.}

\textcolor{black}{Using the negativity of P function {[}\ref{elements of quantum optics}{]},
Lee introduced the criterion for HOA as }

\textcolor{black}{\begin{equation}
R(l,m)=\frac{\left\langle N_{x}^{(l+1)}\right\rangle \left\langle N_{x}^{(m-1)}\right\rangle }{\left\langle N_{x}^{(l)}\right\rangle \left\langle N_{x}^{(m)}\right\rangle }-1<0,\label{eq:ho3}\end{equation}
where $N$ is the usual number operator, $\left\langle N^{(i)}\right\rangle =\left\langle N(N-1)...(N-i+1)\right\rangle $
is the $ith$ factorial moment of number operator, $\left\langle \right\rangle $
denotes the quantum average, $l$ and $m$ are integers satisfying
the conditions $l\leq m\leq1$ and the subscript $x$ denotes a particular
mode. Ba An {[}\ref{ba an}{]} choose $m=1$ and reduced the criterion
of $l$th order antibunching to \begin{equation}
A_{x,l}=\frac{\left\langle N_{x}^{(l+1)}\right\rangle }{\left\langle N_{x}^{(l)}\right\rangle \left\langle N_{x}\right\rangle }-1<0\label{eq:bhuta1}\end{equation}
or, \begin{equation}
\left\langle N_{x}^{(l+1)}\right\rangle <\left\langle N_{x}^{(l)}\right\rangle \left\langle N_{x}\right\rangle .\label{eq:ba an (cond)}\end{equation}
Physically, a state which is antibunched in $l$th order has to be
antibunched in $(l-1)th$ order. Therefore, we can further simplify
(\ref{eq:ba an (cond)}) as \begin{equation}
\left\langle N_{x}^{(l+1)}\right\rangle <\left\langle N_{x}^{(l)}\right\rangle \left\langle N_{x}\right\rangle <\left\langle N_{x}^{(l-1)}\right\rangle \left\langle N_{x}\right\rangle ^{2}<\left\langle N_{x}^{(l-2)}\right\rangle \left\langle N_{x}\right\rangle ^{3}<...<\left\langle N_{x}\right\rangle ^{l+1}\label{eq:ineq}\end{equation}
and obtain the condition for $l-th$ order antibunching as \begin{equation}
d(l)=\left\langle N_{x}^{(l+1)}\right\rangle -\left\langle N_{x}\right\rangle ^{l+1}<0.\label{eq:ho21}\end{equation}
This simplified criterion (\ref{eq:ho21}) coincides exactly with
the physical criterion of HOA introduced by Pathak and Garica {[}\ref{the:martin}{]}.
Here we can note that $d(l)=0$ and $d(l)>0$ corresponds to higher
order coherence and higher order bunching (many photon bunching) respectively.
As we have already mentioned, higher order antibunching is not yet
studied rigourously and apparently, higher order antibunching is very
rare since it is reported only in three particular cases in last 15
years. The present work aims to show that its not really a rare phenomenon
rather it can be seen in many simple optical processes. To establish
that we have used the criterion (\ref{eq:ho21}) and short time approximated
solutions of equation of motions corresponding to various Hamiltonians
(such as six wave mixing process, four wave mixing process and second
harmonic generation) and have shown that HOA can be seen in all the
physical systems selected for the present study. In the next section
we present a second order operator solution of the equation of motion
of six wave mixing process in detail and use that to show the existence
of HOA in six wave mixing process. In section 3 and 4 we have studied
the possibilities of observing higher order antibunching in four wave
mixing process and in second harmonic generation process respectively.
Finally section 5 is dedicated to conclusions. }

\section{Six wave mixing process}

Six wave mixing may happen in different ways. One way is that two
photon of frequency $\omega_{1}$ are absorbed (as pump photon) and
three photon of frequency $\omega_{2}$ and another of frequency $\omega_{3}$
are emitted. The Hamiltonian representing this particular six wave
mixing process is\begin{equation}
H=a^{\dagger}a\omega_{1}+b^{\dagger}b\omega_{2}+c^{\dagger}c\omega_{3}+g(a^{\dagger2}b^{3}c\,+\, a^{2}b^{\dagger3}c^{\dagger})\label{eq:hamilotonian1}\end{equation}
where $a$ and $a^{\dagger}$are creation and annihilation operators
in pump mode which satisfy $[a,a^{\dagger}]$=1, similarly $b,\, b^{\dagger}$
and $c,\, c^{\dagger}$ are creation and annihilation operators in
stokes mode and signal mode respectively and $g$ is the coupling
constant. Substituting $A=a\, e^{i\omega_{1}t},B=b\, e^{i\omega_{2}t}$
and $C=c\, e^{i\omega_{3}t}$ we can write the Hamiltonian (\ref{eq:hamilotonian1})
as \begin{equation}
H=A^{\dagger}A\omega_{1}+B^{\dagger}B\omega_{2}+C^{\dagger}C\omega_{3}+g(A^{\dagger2}B^{3}C\,+\, A^{2}B^{\dagger3}C^{\dagger})\label{eq:hamilotonian}\end{equation}
Since the Hamiltonian is known, we can use Heisenberg's equation of
motion (with $\hbar=1$): \begin{equation}
\dot{A}=\frac{\partial A}{\partial t}+i[H,A]\label{eq:heisenberg}\end{equation}
and short time approximation to find out the time evolution of the
essential operators. From equation (\ref{eq:hamilotonian}) we have
\begin{equation}
\begin{array}{lcl}
[H,A] & = & -A\omega_{1}-2gA^{\dagger}B^{3}C.\end{array}\label{eq:commutation}\end{equation}
From (\ref{eq:heisenberg}) and (\ref{eq:commutation}) we have \begin{equation}
\dot{A}=iA\omega_{1}-iA\omega_{1}-i2gA^{\dagger}B^{3}C=-2igA^{\dagger}B^{3}C.\label{eq:adot}\end{equation}
 Similarly \begin{equation}
\dot{B}=-3igA^{2}B^{\dagger2}C^{\dagger}\label{eq:bdot}\end{equation}
and \begin{equation}
\dot{C}=-igA^{2}B^{\dagger3}.\label{eq:cdot}\end{equation}
We can find the second order differential of $A$ using (\ref{eq:heisenberg})
and (\ref{eq:adot}-\ref{eq:cdot}) as \begin{equation}
\begin{array}{lcl}
\ddot{A} & = & \frac{\partial\dot{A}}{\partial t}+i[H,\dot{A}]\\
 & = & 4g^{2}AB^{\dagger3}B^{3}C^{\dagger}C-18g^{2}A^{\dagger}A^{2}B^{\dagger2}B^{2}C^{\dagger}C-36g^{2}A^{\dagger}A^{2}B^{\dagger}BC^{\dagger}C\\
 & - & 2g^{2}A^{\dagger}A^{2}B^{\dagger3}B^{3}-18g^{2}A^{\dagger}A^{2}B^{\dagger2}B^{2}-36g^{2}A^{\dagger}A^{2}B^{\dagger}B\\
 & - & 12g^{2}A^{\dagger}A^{2}C^{\dagger}C-12g^{2}A^{\dagger}A^{2}.\end{array}\label{eq:adoubledot}\end{equation}
Now by using the Taylor's series expansion \begin{equation}
f(t)=f(0)+t\left(\frac{\partial f(t)}{\partial t}\right)_{t=0}+\frac{t^{2}}{2!}\left(\frac{\partial^{2}f(t)}{\partial t^{2}}\right)_{t=0}......\label{eq:taylor}\end{equation}
and substituting (\ref{eq:adot}) and (\ref{eq:adoubledot}) in (\ref{eq:taylor})
we get 

\begin{equation}
\begin{array}{lcl}
A(t) & = & A-2igtA^{\dagger}B^{3}C\\
 & + & g^{2}t^{2}\left[2AB^{\dagger3}B^{3}C^{\dagger}C-9A^{\dagger}A^{2}B^{\dagger2}B^{2}C^{\dagger}C-18A^{\dagger}A^{2}B^{\dagger}BC^{\dagger}C\right.\\
 & - & \left.A^{\dagger}A^{2}B^{\dagger3}B^{3}-9A^{\dagger}A^{2}B^{\dagger2}B^{2}-18A^{\dagger}A^{2}B^{\dagger}B-6A^{\dagger}A^{2}C^{\dagger}C-6A^{\dagger}A^{2}\right]\end{array}\label{eq:a(t)}\end{equation}
or,\begin{equation}
\begin{array}{lcl}
A(t) & = & A-2igtA^{\dagger}B^{3}C\\
 & + & g^{2}t^{2}\left[2AB^{\dagger3}B^{3}N_{c}-9N_{A}AB^{\dagger2}B^{2}N_{C}-18N_{A}AN_{B}N_{C}\right.\\
 & - & \left.N_{A}AB^{\dagger3}B^{3}-9N_{A}AB^{\dagger2}B^{2}-18N_{A}AN_{B}-6N_{A}AN_{C}-6N_{A}A\right]\end{array}\label{eq:a(t)1}\end{equation}
where $N_{A}=A^{\dagger}A,\:\: N_{B}=B^{\dagger}B,\:\: N_{C}=C^{\dagger}C$%
\footnote{A short time approximated expression of time evolution of annihilation
operator in pump mode of six wave mixing process described by (\ref{eq:hamilotonian1})
is also derived in {[}\ref{giri}{]} but unfortunately their solution
contain some mistakes. %
}. The Taylor series is valid when $t$ is small, so this solution
is valid for a short time and that is why it is called short time
approximation. The above calculation is shown in detail as an example.
Following the same prescription, we can find out the time evolution
of $B$ and $C$ or any other creation and annihilation operator that
appears in the Hamiltonian of matter field interaction. This is a
very strong technique since this straight forward prescription is
valid for any optical process where interaction time is short. After
obtaining the analytic expression for time evolution of annihilation
operator, now we can use it to check whether it satisfies condition
(\ref{eq:ho21}) or not.

Let us start with the possibility of observing first order antibunching.
From equation (\ref{eq:a(t)}), we can derive expressions for $N(t)$
and $N^{(2)}(t)$ as 

\begin{equation}
\begin{array}{lcl}
N(t) & = & A^{\dagger}A-2igt\left(A^{\dagger2}B^{3}C-A^{2}B^{\dagger3}C^{\dagger}\right)\\
 & + & g^{2}t^{2}\left[8A^{\dagger}AB^{\dagger3}B^{3}C^{\dagger}C-18A^{\dagger2}A^{2}B^{\dagger2}B^{2}C^{\dagger}C-36A^{\dagger2}A^{2}B^{\dagger}BC^{\dagger}C-2A^{\dagger2}A^{2}B^{\dagger3}B^{3}\right.\\
 & - & \left.18A^{\dagger2}A^{2}B^{\dagger2}B^{2}-36A^{\dagger2}A^{2}B^{\dagger}B-12A^{\dagger2}A^{2}C^{\dagger}C+4B^{\dagger3}B^{3}C^{\dagger}C-12A^{\dagger2}A^{2}\right]\end{array}\label{eq:N(t)}\end{equation}
and 

\begin{equation}
\begin{array}{lcl}
N^{(2)}(t) & = & A^{\dagger2}(t)A^{2}(t)=A^{\dagger2}A^{2}-2igt\left(2A^{\dagger3}AB^{3}C-A^{\dagger2}B^{3}C-2A^{\dagger}A^{3}B^{\dagger3}C^{\dagger}-A^{2}B^{\dagger3}C^{\dagger}\right)\\
 & + & g^{2}t^{2}\left[24A^{\dagger2}A^{2}B^{\dagger3}B^{3}C^{\dagger}C+32A^{\dagger}AB^{\dagger3}B^{3}C^{\dagger}C+4B^{\dagger3}B^{3}C^{\dagger}C\right.\\
 & - & 36A^{\dagger3}A^{3}B^{\dagger2}B^{2}C^{\dagger}C-72A^{\dagger3}A^{3}B^{\dagger}BC^{\dagger}C-18A^{\dagger2}A^{2}B^{\dagger2}B^{2}C^{\dagger}C\\
 & - & 36A^{\dagger2}A^{2}B^{\dagger}BC^{\dagger}C-4A^{\dagger3}A^{3}B^{\dagger3}B^{3}-36A^{\dagger3}A^{3}B^{\dagger2}B^{2}\\
 & - & 72A^{\dagger3}A^{3}B^{\dagger}B-2A^{\dagger2}A^{2}B^{\dagger3}B^{3}-18A^{\dagger2}A^{2}B^{\dagger2}B^{2}\\
 & - & 4A^{\dagger4}B^{6}C^{2}-4A^{4}B^{\dagger6}C^{\dagger2}-36A^{\dagger2}A^{2}B^{\dagger}B\\
 & - & \left.24A^{\dagger3}A^{3}C^{\dagger}C-12A^{\dagger2}A^{2}C^{\dagger}C-24A^{\dagger3}A^{3}-12A^{\dagger2}A^{2}\right].\end{array}\label{eq:Nsquare(t)}\end{equation}

In the present study, we have taken all the expectations with respect
to $|\alpha>|0>|0>$ for simplification. This assumption physically
means that initially a coherent state (say, a laser) is used as pump
and before the interaction of the pump with atom, there was no photon
in signal mode ($b$) or stokes mode ($c$). Thus the pump interacts
with the atom and causes excitation followed by emissions. Now from
(\ref{eq:N(t)}) and (\ref{eq:Nsquare(t)}), we have

\begin{equation}
\left\langle N(t)\right\rangle ^{2}=|\alpha|^{4}-24g^{2}t^{2}|\alpha|^{6}.\label{eq:Nexpectationsquare}\end{equation}

\begin{equation}
\left\langle N^{2}(t)\right\rangle =|\alpha|^{4}-g^{2}t^{2}\left(24|\alpha|^{6}+12|\alpha|^{4}\right)\label{eq:n2}\end{equation}
where $A|\alpha>=\alpha|\alpha>$. Now by using (\ref{eq:Nexpectationsquare})
and (\ref{eq:n2}) we can show that the six wave mixing process satisfies
the criterion of antibunching (\ref{eq:ho21}) since:

\begin{equation}
\begin{array}{lcl}
d(1) & = & \left\langle N^{(2)}(t)\right\rangle -\left\langle N(t)\right\rangle ^{2}\\
 & = & \left[|\alpha|^{4}+g^{2}t^{2}(-24|\alpha|^{6}-12|\alpha|^{4})\right]-\left[|\alpha|^{4}-24g^{2}t^{2}|\alpha|^{6}\right]\\
 & = & -12g^{2}t^{2}|\alpha|^{4}.\end{array}\label{eq:d(1)}\end{equation}
From the last equation it is clear that $d(1)$ is always negative,
i.e. it always shows usual antibunching. Essentially, this is a nonclassical
state but mere satisfaction of nonclassicality or antibunching is
not enough because we are looking for HOA. Let us see what happens
in the next higher order.

For the study of second order of antibunching, $A^{3}(t)$ can be
obtained by using {[}\ref{eq:a(t)}{]} and operator ordering techniques:

\begin{equation}
\begin{array}{lcl}
A^{3}(t) & = & A^{3}-2igt\left(3A^{\dagger}A^{2}B^{3}C+3AB^{3}C\right)\\
 & + & g^{2}t^{2}\left[6A^{3}B^{\dagger3}B^{3}C^{\dagger}C-27A^{\dagger}A^{4}B^{\dagger2}B^{2}C^{\dagger}C-54A^{\dagger}A^{4}B^{\dagger}BC^{\dagger}C\right.\\
 & - & 54A^{3}B^{\dagger}BC^{\dagger}C-27A^{3}B^{\dagger2}B^{2}C^{\dagger}C-3A^{\dagger}A^{4}B^{\dagger3}B^{3}-27A^{\dagger}A^{4}B^{\dagger2}B^{2}-54A^{\dagger}A^{4}B^{\dagger}B\\
 & - & 3A^{3}B^{\dagger3}B^{3}-27A^{3}B^{\dagger2}B^{2}-18A^{\dagger}A^{4}C^{\dagger}C-18A^{3}C^{\dagger}C\\
 & - & \left.54A^{3}B^{\dagger}B-12A^{\dagger2}AB^{6}C^{2}-12A^{\dagger}B^{6}C^{2}-18A^{\dagger}A^{4}-18A^{3}\right]\end{array}\label{eq:acube(t)}\end{equation}
Then $A^{\dagger3}(t)$ can simply be written as,\begin{equation}
\begin{array}{lcl}
A^{\dagger3}(t) & = & A^{\dagger3}+2igt\left(3A^{\dagger2}AB^{\dagger3}C^{\dagger}+3A^{\dagger}B^{\dagger3}C^{\dagger}\right)\\
 & + & g^{2}t^{2}\left[6A^{\dagger3}B^{\dagger3}B^{3}C^{\dagger}C-27A^{\dagger4}AB^{\dagger2}B^{2}C^{\dagger}C-54A^{\dagger4}AB^{\dagger}BC^{\dagger}C\right.\\
 & - & 54A^{\dagger3}B^{\dagger}BC^{\dagger}C-27A^{\dagger3}B^{\dagger2}B^{2}C^{\dagger}C-3A^{\dagger4}AB^{\dagger3}B^{3}-27A^{\dagger4}AB^{\dagger2}B^{2}-54A^{\dagger4}AB^{\dagger}B\\
 & - & 3A^{\dagger3}B^{\dagger3}B^{3}-27A^{\dagger3}B^{\dagger2}B^{2}-18A^{\dagger4}AC^{\dagger}C-18A^{\dagger3}C^{\dagger}C\\
 & - & \left.54A^{\dagger3}B^{\dagger}B-12A^{\dagger}A^{2}B^{\dagger6}C^{\dagger2}-12AB^{\dagger6}C^{\dagger2}-18A^{\dagger4}A-18A^{\dagger3}\right]\end{array}\label{eq:Adaggercube(t)}\end{equation}
Last two equations can be used to calculate the third factorial moment
$(N^{(3)}(t))$ of the number operator $N$ as

\begin{equation}
\begin{array}{lcl}
N^{(3)}(t) & = & A^{\dagger3}A^{3}-2igt\left(3A^{\dagger4}A^{2}B^{3}C+3A^{\dagger3}AB^{3}C-3A^{\dagger2}A^{4}B^{\dagger3}C^{\dagger}-3A^{\dagger}A^{3}B^{\dagger3}C^{\dagger}\right)\\
 & + & g^{2}t^{2}\left[48A^{\dagger3}A^{3}B^{\dagger3}B^{3}C^{\dagger}C-54A^{\dagger4}A^{4}B^{\dagger2}B^{2}C^{\dagger}C-54A^{\dagger3}A^{3}B^{\dagger2}B^{2}C^{\dagger}C\right.\\
 & - & 108A^{\dagger3}A^{3}B^{\dagger}BC^{\dagger}C+108A^{\dagger2}A^{2}B^{\dagger3}B^{3}C^{\dagger}C+36A^{\dagger}AB^{\dagger3}B^{3}C^{\dagger}C\\
 & - & 108A^{\dagger4}A^{4}B^{\dagger}BC^{\dagger}C-54A^{\dagger4}A^{4}B^{\dagger2}B^{2}-108A^{\dagger4}A^{4}B^{\dagger}B-6A^{\dagger3}A^{3}B^{\dagger3}B^{3}\\
 & - & 6A^{\dagger4}A^{4}B^{\dagger3}B^{3}-54A^{\dagger3}A^{3}B^{\dagger2}B^{2}-108A^{\dagger3}A^{3}B^{\dagger}B-12A^{\dagger5}AB^{6}C^{2}-12A^{\dagger}A^{5}B^{\dagger6}C^{\dagger2}\\
 & - & \left.12A^{\dagger4}B^{6}C^{2}-12A^{4}B^{\dagger6}C^{\dagger2}-36A^{\dagger3}A^{3}C^{\dagger}C-36A^{\dagger4}A^{4}C^{\dagger}C-36A^{\dagger4}A^{4}-36A^{\dagger3}A^{3}\right].\end{array}\label{eq:Ncube(t)}\end{equation}
Taking the expectation value with respect to the intial state we obtain

\begin{equation}
\begin{array}{lcl}
\left\langle N^{(3)}(t)\right\rangle  & = & |\alpha|^{6}-g^{2}t^{2}\left(36|\alpha|^{8}+36|\alpha|^{6}\right).\end{array}\label{eq:bhuta2}\end{equation}
On the other hand, 

\begin{equation}
\left\langle N(t)\right\rangle ^{3}=|\alpha|^{6}-36g^{2}t^{2}|\alpha|^{8}\label{eq:Nexpectationcube}\end{equation}
By using the last two equations we can see that the pump mode of six
wave mixing process satisfy the criteria of antibunching of second
order (\ref{eq:ho21}). Since,

\begin{equation}
\begin{array}{lcl}
d(2) & = & \left\langle N^{(3)}(t)\right\rangle -\left\langle N(t)\right\rangle ^{3}\\
 & = & \left[|\alpha|^{6}-g^{2}t^{2}(36|\alpha|^{8}+36|\alpha|^{6})\right]-\left[|\alpha|^{6}-36g^{2}t^{2}|\alpha|^{8}\right]\\
 & = & -36g^{2}t^{2}|\alpha|^{6}\end{array}\label{eq:d(2)}\end{equation}
is always negative.

\section{Four wave mixing process}

Similarly, four wave mixing may also happen in different ways. One
way is that in which two photon of frequency $\omega_{1}$ are absorbed
(as pump photon) and one photon of frequency $\omega_{2}$ and another
of frequency $\omega_{3}$ are emitted. The Hamiltonian representing
this particular four wave mixing process is

\begin{equation}
H=a^{\dagger}a\omega_{1}+b^{\dagger}b\omega_{2}+c^{\dagger}c\omega_{3}+g(a^{\dagger2}bc\,+\, a^{2}b^{\dagger}c^{\dagger}).\label{eq:hfourwave}\end{equation}
 Following the same prescription as it is used in six wave case we
can write the solution as

\begin{equation}
A(t)=A-2igtA^{\dagger}BC+\frac{g^{2}t^{2}}{2!}[4AB^{\dagger}BC^{\dagger}C-2A^{\dagger}A^{2}B^{\dagger}B-2A^{\dagger}A^{2}C^{\dagger}C-2A^{\dagger}A^{2}]\label{eq:afourwave}\end{equation}
 or 

\begin{equation}
A(t)=A-2igtA^{\dagger}BC+g^{2}t^{2}[2AN_{B}N_{c}-N_{A}AN_{B}-N_{A}AN_{C}-N_{A}A].\label{eq:a-fourwave1}\end{equation}
The respective values of the first order antibunching ($d(1)$) and
second order antibunching ($d(2)$) can similarly be calculated as
we have done for six wave mixing and that yields

\begin{equation}
\begin{array}{lcl}
d(1) & = & \left\langle N^{(2)}(t)\right\rangle -\left\langle N(t)\right\rangle ^{2}\\
 & = & \left[|\alpha|^{4}+g^{2}t^{2}(-4|\alpha|^{6}-2|\alpha|^{4})\right]-\left[|\alpha|^{4}-4g^{2}t^{2}|\alpha|^{6}\right]\\
 & = & -2g^{2}t^{2}|\alpha|^{4}.\end{array}\label{eq:di-fourwave}\end{equation}
 and 

\begin{equation}
\begin{array}{lcl}
d(2) & = & \left\langle N^{(3)}(t)\right\rangle -\left\langle N(t)\right\rangle ^{3}\\
 & = & \left[|\alpha|^{6}-g^{2}t^{2}(6|\alpha|^{8}+6|\alpha|^{6})\right]-\left[|\alpha|^{6}-6g^{2}t^{2}|\alpha|^{8}\right]\\
 & = & -6g^{2}t^{2}|\alpha|^{6}\end{array}\label{eq:d(2)-fourwave}\end{equation}
which are negative and thus they satisfy our criterion for antibunching
and HOA respectively.

\section{Second harmonic generation}

The similar procedure can be repeated for the second harmonic generation
process whose Hamiltonian is 

\begin{equation}
H=\hbar\omega N_{1}+2\hbar\omega N_{2}+hg\left(a_{2}^{\dagger}a_{1}^{2}+a_{1}^{\dagger2}a_{2}\right).\label{eq:hseconh}\end{equation}
The second order expression of the time evolution of annihilation
operator in pump mode of second harmonic generation is

\begin{equation}
A(t)=a_{1}-2igta_{1}^{\dagger}a_{2}+2g^{2}t^{2}\left(a_{2}^{\dagger}a_{2}a_{1}-\frac{1}{2}a_{1}^{\dagger}a_{1}^{2}\right).\label{eq:asecondh}\end{equation}
 Using the last equation along with the method used in section 2 we
obtain

\begin{equation}
\begin{array}{lcl}
d(1) & = & \left\langle N^{(2)}(t)\right\rangle -\left\langle N(t)\right\rangle ^{2}\\
 & = & \left[|\alpha|^{4}+g^{2}t^{2}(-4|\alpha|^{6}-2|\alpha|^{4})\right]-\left[|\alpha|^{4}-4g^{2}t^{2}|\alpha|^{6}\right]\\
 & = & -2g^{2}t^{2}|\alpha|^{4}\end{array}\label{eq:d1-secondh}\end{equation}
and

\begin{equation}
\begin{array}{lcl}
d(2) & = & \left\langle N^{(3)}(t)\right\rangle -\left\langle N(t)\right\rangle ^{3}\\
 & = & \left[|\alpha|^{6}-g^{2}t^{2}(6|\alpha|^{8}+6|\alpha|^{6})\right]-\left[|\alpha|^{6}-6g^{2}t^{2}|\alpha|^{8}\right]\\
 & = & -6g^{2}t^{2}|\alpha|^{6}\end{array}\label{eq:d(2) higher harmonic}\end{equation}
 which are both negative and hence satisfies the criterion (\ref{eq:ho21})
for antibunching and HOA respectively.

\section{Conclusions:}

\textcolor{black}{From (\ref{eq:d(2)}, \ref{eq:d(2)-fourwave} and
\ref{eq:d(2) higher harmonic}), it is clear that all the physical
systems selected for the present study show second order antibunching
i.e. higher order sub-poissonian photon statistics. Thus the present
work strongly establishes the fact that HOA is not a rare phenomenon.
The physical systems studied in the present paper are simple and easily
achievable in laboratories and thus it opens up the possibility of
experimental observation of HOA. In case of interaction of intense
electromagnetic field with third order nonlinear medium it was reported
{[}\ref{the:martin}{]} that the degree of antibunching ($d(l)$)
can be tuned because they depend strongly on the phase of the input
field which can be tuned. This is not the case with any of the physical
systems studied in the present work. It is also clear that the higher
order antibunching would not have been observed if we would have considered
first order operator solutions (first order in $g$), on the other
hand, if we use second order operator solutions then the depth of
nonclassicality is found to increase monotonically with the increase
of input photon number ($|\alpha|^{2}$). Possibly this monotonic
increment will be ceased by the higher order perturbation terms. It
is also observed that if we assume that the anharmonic constant and
number of photon initially present in the pump mode are same for all
three cases then the depth of nonclassicality is same in four wave
mixing and second harmonic generation process and its more in six
wave mixing process. }

\textcolor{black}{The prescription followed in the present work is
easy and straight forward and it can be used to study the possibilities
of observing higher order antibunching in other physical systems.
Thus it opens up the possibility of studying higher order nonclassical
effects from a new perspective. This is also important from the application
point of view because any probabilistic single photon source used
for quantum cryptography has to satisfy the condition for higher order
antibunching. Therefore, the simple prescription followed in the present
work may help us to compare the existing sources of single photon. }

\end{document}